\documentstyle[12pt]{article}
\begin{document}

\newtheorem{theorem}{Theorem}
\def\ep{\varepsilon}
\newcommand{\ar}{\rightarrow}
\def\re{{\rm Re}}
\def\im{{\rm Im}}
\def\const{{\rm const.}}
\def\vp{\varphi}
\def\om{\omega}
\def\d{\delta}
\def\a{\alpha}
\def\b{\beta}
\def\l{\lambda}
\def\R{\mbox{\bf R}}
\def\T{\mbox{\bf T}}
\def\wvp{\widehat{\varphi}}

\noindent{\Large\sf Global existence and exponential decay for
hyperbolic dissipative relativistic fluid theories} 

\vspace{.3cm}

{\sf Heinz-Otto Kreiss}

{\sl Department of Mathematics, The University of California, Los
Angeles, CA 

90024, USA.}

\vspace{3truemm}

{\sf Gabriel B. Nagy, Omar E. Ortiz, and Oscar A.
Reula\footnote{Researcher of  CONICET}}

{\sl Fa.M.A.F. - U.N.C., Dr. Medina Allende y Haya de la Torre, Ciudad
Universi-

taria, (5000) C\'ordoba, Argentina.}

\begin{abstract} 
We consider dissipative relativistic fluid theories on a fixed flat,
globally hyperbolic, Lorentzian manifold $(\R \times {\bf
T}^3,g_{ab})$.  We prove that for all initial data in a small enough
neighborhood of the equilibrium states (in an appropriate Sobolev
norm), the solutions evolve smoothly in time forever and decay
exponentially to some, in general undetermined, equilibrium state.  To
prove this, three conditions are imposed on these theories. The first
condition requires the system of equations to be symmetric hyperbolic,
a fundamental requisite to have a well posed and physically consistent
initial value formulation. The second condition is a generic
consequence of the entropy law, and is imposed on the non principal
part of the equations. The third condition is imposed on the principal
part of the equations and it implies that the dissipation affects all
the fields of the theory. With these requirements we prove that all the
eigenvalues of the symbol associated to the system of equations of the
fluid theory have strictly negative real parts, which in fact, is an
alternative characterization for the theory to be totally dissipative.
Once this result has been obtained, a straight forward application of a
general stability theorem due to Kreiss, Ortiz, and Reula, implies the
results above mentioned.
\end{abstract}

\newpage

\section*{\normalsize \sf I. INTRODUCTION}

In recent years there has been a substantial improvement on our
understanding on how a proper description of dissipative fluids can be
incorporated in the framework of the theory of relativity. Dissipative
relativistic fluid theories satisfying an entropy law, and having a
well posed (symmetric hyperbolic), and causal initial value formulation
have been presented \cite{lmr,gl1,gl2}.

An important result on the physical meaning of all these hyperbolic
theories was obtained recently \cite{g,l}. It was shown that
certain constitutive relations between the variables in the hyperbolic
fluid theories, which have a clear physical meaning, approach in their
time evolution the values predicted by the simplest covariant
generalizations  of the Navier Stokes fields. This result is based in a
fundamental hypothesis namely, that the solution of the hyperbolic
fluid field equations exists and remains smooth and small during a
long enough time interval, such that the relaxation to near
Navier-Stokes behavior occurs.  This is a very important check for
these fluid theories, since at microscopic scales they are
substantially different from the usual Navier-Stokes theories in the
following sense: for the last ones, one expects to have smooth global
solutions for all smooth initial data sets, as has been proved in
lower dimensions \cite{kl}, while for the former ones, one expects the
developing of discontinuities in the form of shock waves for crispy
enough initial data.  Thus, we can only hope to find a neighborhood of
equilibrium data for which global solutions exist and so, where the
departures from Navier-Stokes are uniformly small.

The purpose of the present work is to look for conditions under which
the fundamental hypothesis mentioned above is satisfied. To this end,
we apply a theorem \cite{kor} which is a generalization to the case of
partial differential equations of the Ljapunov stability theorem for
ordinary differential equations. This generalization holds for
hyperbolic systems such that the eigenvalues of their associated symbols
have all strictly negative real parts which, as we shall see, is the
case for the hyperbolic dissipative fluids.

In order to apply this general stability theorem, three conditions are
imposed on the dissipative hyperbolic fluid theories.  The first
condition requires the symmetric hyperbolicity of the system of
equations; a fundamental requisite to have a well posed and physically
consistent initial value formulation. As we shall see, the symmetry is
an automatic property of these theories, but the hyperbolicity has to
be required. We also include in this condition that the spacetime
manifold is $\R\times \T^3$ with a flat metric (which is the more
restrictive assumption). The other two conditions are of a generic type
in the sense that all fluid systems,
except for very specific and isolated ones, satisfy them.
Specifically, the second condition requires that the nonprincipal part
of the system of equations, which is responsible for the dissipation,
satisfy certain negative-semidefiniteness condition. This condition
assures that all perturbations to an equilibrium state, which are not
tangent to the equilibrium submanifold, do dissipate towards
equilibrium, and this is manifested by the fact that they make a
positive definite contribution to the entropy. The third condition is
a requirement 
on the principal part of the system of equations, and it means that the
presence of dissipation affects all the fields of the theory, in the
sense of not allowing for a decoupled set of fields with its own
evolution not being driven by dissipation. Both, the second and the
third conditions, have already been required in the literature with the
aim of characterizing the equilibrium states.

These conditions allow us to apply the theorem proved in \cite{kor},
which implies not only the global existence of solutions, but also
their exponential decay to equilibrium for initial data near enough, in
some appropriate norm, to equilibrium data.

The plan of the paper is as follows: In Sec. II we briefly introduce
the fundamental aspects of the fluid theories, and state in detail the
character of the conditions we impose on them.  In Sec. III we state
and prove our main result. Finally, in Sec. IV we present the
conclusions.

\section*{\normalsize \sf II. DISSIPATIVE RELATIVISTIC FLUID THEORIES}

In this section, following \cite{gl2}, we introduce the dissipative
relativistic fluid theories. After introducing them, we describe the
properties of these fluids needed to prove stability.

We assume that a fluid state is characterized by a finite collection of
space-time tensor fields. Let $\varphi^A$ denote these fields, where
upper case indices stand for the entire set of tensor indices
represented in this collection of fields. So we refer to $\varphi^A$ as
a point in the space of fluid states ${\cal S}$. Lower
case indices will denote space-time indices. Repeated indices indicate contraction
as usual in the abstract index notation. We restrict consideration to
the fluid theories in which the field equations take the form
\begin{equation}
M^a_{AB}\nabla_a\varphi^B = -I_{AB} \vp^B \label{sistgen}.
\end{equation} 
where $M^a_{AB}$ and $I_{AB}$ are smooth functions of the fluid fields
$\vp^A$ and the space-time metric. 

We say that system (\ref{sistgen}) is symmetric hyperbolic if
$M^a_{AB}=M^a_{(AB)}$ and there exists a timelike-future directed $u^a$
such that $N_{AB} \equiv -u_aM^a_{AB}$ is positive definite. This is a
sufficient condition to have a well posed initial value formulation.

On physical grounds, to have as maximum speed of propagation the speed
of light, the stronger condition of causality, that is, $-u_aM^a_{AB}$
is positive definite for all future-directed timelike vector $u^a$, is
usually required. This means that the characteristic surfaces of system
(\ref{sistgen}) are inside the null cone given by the space-time
metric. To prove global existence and decay only the weaker condition
of hyperbolicity is needed.

We analize now the structure of the equilibrium states. Following
\cite{gl1,gl2} we say that a fluid state $\varphi^A$, solution of the
dynamical system of equations, is a {\em strict equilibrium state} if
its time reverse is also a solution. We denote a strict equilibrium
state by $\varphi^A_0$, and assign a subindex zero to any tensor
evaluated at an strict equilibrium state. This definition implies that
$I_0{}_{AB} \vp^B_0=0$. More generally, we say the $\vp^A$ is a {\em
momentary equilibrium state} if $I_{AB} \vp^B=0$. These states are
called equilibrium states because their entropy production vanishes,
and momentary because if this condition holds at a certain time, it
will not necessarily hold in subsequent times (see \cite{g}). Every
fluid state can be written as $\vp^A=\psi^A+\eta^A$, where $I_{AB}
\psi^B=0$, and $\eta^A$ is such that $I_{AB}\eta^B=0$ implies
$\eta^A=0$. We call $\psi^A$ the momentary equilibrium part of the
fluid state.

Below we state the three conditions imposed on these fluid theories
that will be used to show global existence and exponential decay.

\begin{enumerate}

\item The fluid system of equations (\ref{sistgen}) is symmetric
hyperbolic and the space-time is $(\R \times {\bf T}^3, g_{ab})$, where
${\bf T}^3$ denotes a three dimensional torus and $g_{ab}$ is a flat
metric.

\item The tensor $I_{0AB}$ must be symmetric and positive
semidefinite.

\item The map ${\cal F}_K: {\cal S}_{\psi} \to {\cal S}^*$ defined by
${\cal F}_K(\psi^A) \equiv K_a M^a_0{}_{AB}\; \psi^B$, is injective for
all space-time vectors $K^a \neq 0$; where ${\cal S}^*$ denotes the
dual of the space of fluid states and ${\cal S}_{\psi}$ the subspace of
momentary equilibrium states.

\end{enumerate}
The first condition is the more restrictive and more work has to be
done in order to weaken it. It would be interesting to treat physically
relevant boundary conditions and arrive to similar results. The other
two conditions are not very restrictive and all fluid systems, except
for very specific and isolated ones, satisfy them.

The second requirement ensures that the effect of $I_{AB}$ in equation
(\ref{sistgen}) is to dissipate, in the sense of tending to move non
equilibrium states towards equilibrium as time grows. This is just a
little stronger than the entropy condition in these fluid theories,
that requires the entropy source to be non negative. This stronger
condition was already considered in \cite{gl1}.

The third requirement is on the principal part of the equations. It
implies that dissipation affects all the fields of the theory, in the
sense of not allowing the existence of a decoupled subset of fields
with its own evolution not being affected by dissipation. This
requirement turns out to be equivalent (see appendix A), at least for
the case of divergence type fluid theories, to one assumed in the
literature \cite{gl1} to show that the only possible strict equilibrium
states are the constant states \cite{equilibrium}.

\section*{\normalsize \sf III. GLOBAL EXISTENCE AND EXPONENTIAL DECAY}   

In the theorem below we present our main result about global existence
in time and decay to strict equilibrium.

\begin{theorem}\label{teo}
Consider the Cauchy problem for system (\ref{sistgen}), corresponding
to a hyperbolic divergence type fluid theory satisfying conditions
1-3.  If the initial data is smooth and close enough in a $H^p(\T^3)$ Sobolev norm
($p>5$) to the data corresponding to some strict equilibrium solution,
then the solution is smooth, exists globally in time and decays exponentially to
some strict equilibrium solution in the $H^p$ norm.
\end{theorem}

{\it Proof:} We define $N_{AB} = -u_a M^a_{AB}$, $N^a_{AB} =
-q^a_b M^b_{AB}$ where $q_{ab}= g_{ab} + u_a u_b$ is the 3-metric in
each hypersurface orthogonal to $u^a$ (assuming $u^a u_a =-1$). Assume
that the initial data is close to some strict equilibrium state
$\vp_0^A$, then the fluid state is $\vp^A = \vp_0^A + \d \vp^A$.  For
the variable $\d\vp^A$, the system (\ref{sistgen}) becomes
\begin{equation}
N_{AB} u^a \nabla_a \d\vp^B = N^a_{AB} \nabla_a \d\varphi^B - I_{AB}
\d\vp^B. \label{sistf}
\end{equation}
Here the tensors $N_{AB}$, $N^a_{AB}$, and $I_{AB}$ are thought as
functions of $\d\vp^A$. The condition (1) of Sec. II on the fluid
theories implies that we can choose Cartesian
coordinates $\{t,x^j\}$ on the space-time manifold $\R \times {\bf
T}^3$ and a constant $u^a$ such that $\partial/\partial t =u^a
\nabla_a$.  Then the system (\ref{sistf}) becomes
\begin{equation}
N_{AB} \frac{\partial}{\partial t}\d\vp^B = N^j_{AB} \frac{\partial
\d\vp^B}{\partial x^j} - I_{AB} \d\vp^B. \label{sistf1}
\end{equation}
In order to study solutions near $\d\vp^A=0$, it is convenient to
introduce a parameter $\ep$ to control the smallness of initial data.
Thus, $\d\vp^A(t=0) = \ep f^A(x^j)$ and the solution shall be written
as $\d\vp^A = \ep v^A$. As the tensors $N^j_{AB}$ and $I_{AB}$ are
smooth functions of $\ep v^A$, they can be written as
\begin{eqnarray*}
&&N^j_{AB} = N^j_{0AB} + \ep N^j_{1AB} ,\\
&&I_{AB} = I_{0AB} + \ep I_{1AB}.
\end{eqnarray*}
With this decomposition the Cauchy problem for (\ref{sistf1}) is
\begin{eqnarray}
&&N_{AB} \frac{\partial v^B}{\partial t} = \Bigl(N^j_{0AB} + \ep
N^j_{1AB}\Bigr) \frac{\partial v^B}{\partial x^j} + \Bigl(I_{0AB} + \ep
I_{1AB}\Bigr) v^B. \label{sistf2} \\
&& v^A(t=0)=f^A(x^j). \nonumber
\end{eqnarray}
As there are periodic boundary conditions on the space coordinates
$\{x^j\}$, $v^A$ can be expanded in Fourier series,
$$
v^A = \sum_{k^j \in \Omega} \hat v^A(k^j,t) \; e^{i \vec{x}\cdot\vec{k}},
$$
where $\Omega$ is the discrete set of Fourier frequencies.

We want to apply the stability theorem proved in \cite{kor} that, for
completeness, we state in the appendix B. To prove the theorem \ref{teo},
we consider the eigenvalues problem
\begin{equation}
\lambda N_{0AB} \wvp^B = \bigl(i k_j N^j_{0AB} - I_{0AB}\bigr)\wvp^B.
\label{eigpro}
\end{equation}
Then, as explained in the appendix B, we only need to verify the
following conditions.

(i) There is a constant $\d>0$ such that the eigenvalues $\lambda(k^j)$
satisfy $\re\{\l\} \leq -\d$ for all $k^j \in \Omega$, $k^j \neq 0$.

(ii) For $k^j=0$, $\l(0)\leq -\d$ or $\l(0)=0$. 

(iii) As linear maps, the null space of $I_{1AB}$ contains the null space of
$I_{0AB}$.

\noindent Conditions (ii) and (iii) are satisfied for these fluid
theories, since the kernel of $I_{AB}$ is of constant dimension (see
\cite{gl1,gl2}), and because condition (2) of Sec. II holds.

To prove (i), let $k^j$ be different from zero, then $k=\sqrt{k^j
k_j}\geq {\rm const.}>0$. The eigenvalue problem (\ref{eigpro}) can be
written as
$$
\Bigl(-\frac{\l}{k} N_{0AB} + i\frac{k_j}{k} N^j_{0AB} \Bigr)
\widehat{\psi}^B = \Bigl(\frac{\l}{k} N_{0AB} + \frac{1}{k} I_{0AB} -
i\frac{k_j}{k} N^j_{0AB} \Bigr) \widehat{\eta}^B.
$$
Defining $K_a= -(\l/k) u_a + i (k_j/k) q^j_a$, this can be written as 
\begin{equation}
K_a M^a_{0AB} \widehat{\psi}^B =\Bigl(\frac{\l}{k} N_{0AB}
+ \frac{1}{k} I_{0AB} - i\frac{k_j}{k} N^j_{0AB} \Bigr)
\widehat{\eta}^B. \label{epd}
\end{equation}
Since $K_a M^a_{0AB}$ is injective, by condition (3), and a smooth
function of $k_j/k$, there is a constant $c>0$ such that
$$
c \; N_{0BC}  \overline{\widehat{\psi}}^B \widehat{\psi}^C \leq K_a
{{M^a_0}^A}_B K_b M^b_{0AC}  \overline{\widehat{\psi}}^B
\widehat{\psi}^C.
$$
Then, contracting (\ref{epd}) with itself
\begin{eqnarray*}
&& N_{0BC} \overline{\widehat{\psi}}^B \widehat{\psi}^C \leq \\
&& \frac{1}{c} \Bigl(\frac{\overline{\l}}{k} {{N_0}^A}_B + \frac{1}{k}
{{I_0}^A}_B + i\frac{k_j}{k} {{N^j_0}^A}_B \Bigr)
\Bigl(\frac{\l}{k} N_{0AC} + \frac{1}{k} I_{0AC} - i\frac{k_j}{k}
N^j_{0AC} \Bigr) \overline{\widehat{\eta}}^B \widehat{\eta}^C.
\end{eqnarray*}
Perturbation theory of linear operators tells us that $\l(k^j)/k$ is
uniformly bounded (see \cite{k}), then we get for some positive $c'$
$$
N_{0BC} \overline{\widehat{\psi}}^B \widehat{\psi}^C \leq c' \; 
N_{0BC} \overline{\widehat{\eta}}^B \widehat{\eta}^C. \label{ievp}
$$
This inequality, together with the positive definiteness of $N_{0AB}$
implies,
\begin{equation}
N_{0AB} \overline{\wvp}^A \wvp^B \leq \tilde c \;  N_{0AB}
\overline{\widehat{\eta}}^A \widehat{\eta}^B, \hspace{1cm} \tilde c>0.
\label{iev}
\end{equation}

Now, contracting (\ref{eigpro}) with $\overline{\wvp}^A$ and taking
real part
\begin{eqnarray*}
\re\{\l\} N_{0AB}  \overline{\wvp}^A \wvp^B &=& -I_{0AB}
\overline{\wvp}^A \wvp^B = -I_{0AB} \overline{\widehat{\eta}}^A
\widehat{\eta}^B \\
&\leq& -\tilde \d \; N_{0AB} \overline{\widehat{\eta}}^A
\widehat{\eta}^B \\
&\leq& -\frac{\tilde \d}{\tilde c} \; N_{0AB} \overline{\wvp}^A \wvp^B.
\label{desi}
\end{eqnarray*}
Here, $\tilde \d>0$ exists because of the negative definiteness of
$I_{0AB}$ in the direction of $\widehat{\eta}^A$ (by condition (2)),
and we have used (\ref{iev}) in the last line. We have thus shown
that
$$
\re\{\l(k^j)\} \leq -\d <0, \hspace{1cm} \mbox{with } \d=\frac{\tilde
\d}{\tilde c}>0 \; \mbox{ and } k^j \neq 0.
$$

\section*{\normalsize \sf IV. CONCLUSIONS}

In this work we have proved global existence and exponential decay to
strict equilibrium of solutions, corresponding to initial data in a
small enough neighborhood of strict equilibrium, for a generic
dissipative relativistic fluid theory.

This result, in particular,  verifies a fundamental hypothesis of
previous works \cite{g,l} namely, the existence of solutions during a
long enough time interval. The closeness of initial data to strict
equilibrium is a natural limitation in the sense that, for large data,
shock waves develop, which are a widely observed phenomena in
nature. This occurs because these fluid systems are genuinely
non-linear.

There are three questions related to the techniques used in this work
and the possibility of improving on them. One is whether it is possible
to extend the present result, or rather the general theorem used, to
the case of non-constant equilibrium states, this is of vital
importance if we want to consider non-flat backgrounds or even
self-gravitating fluids. We think this is probable the case if we
further assume that the theory has, at equilibrium, a conserved
energy--that is a positive definite bilinear form in the tangent space
of equilibrium states--as is usually the case for theories coming from
a Hamiltonian formalism. The second possible extension is towards
allowing non-compact Cauchy slices. There is another technique,
introduced by Matsumura \cite{m}, which allows to study global
existence and stability for some particular cases that ranges from
hyperbolic heat conduction to relativistic superfluids \cite{kor1}.
This technique allows to treat the case of non-compact Cauchy slices,
but can not be applied to the general systems considered in this work.
Thus, it would be important to extend the theorem given in the appendix
B to the case of non-compact Cauchy slices. The third extension is in
the direction of boundary values problems. It is clear that one would
like to use this theory to describe situations where the fluid is in a
finite region of space, in that case the equations cease to be
hyperbolic outside the region occupied by the fluid, and so a boundary
value formulation is needed.  Since Navier-Stokes fluids behave in a
much more amenable way with regards to boundary conditions than perfect
fluids, one would expect that these dissipative fluids will have that
property too, making this an interesting, and perhaps tractable,
problem.

\section*{\normalsize \sf ACKNOWLEDGMENT}

G.B.N., O.E.O. and O.A.R. thank ICTP for hospitality during part of the time
of development of this work.

This work has been supported by grants from CONICOR, CONICET and
Se\-CyT-UNC. 

\section*{\normalsize \sf APPENDIX A: EQUIVALENCE OF ASSUMPTIONS}

In this Appendix we prove the equivalence between the requirement (3)
of Section 2 and a condition imposed in literature \cite{gl1,gl2}. We
do this, for simplicity, in the case of divergence type fluid theories,
and we assume that the reader is familiar with \cite{gl1}.  The
condition under consideration allows to characterize the strict
equilibrium states in such a way that they are precisely the same set
of equilibrium states found for the standard Eckart theory.

The requirement 3 of Section 2 is the following: The map ${\cal F}_K:
{\cal S}_{\psi} \to {\cal S}^*$ defined by ${\cal F}_K ( \psi^A )
\equiv K_a M^a_0{}_{AB} \; \psi^B$, is injective for all space-time
vectors $K^a \neq 0$, where ${\cal S}^*$ denotes the dual of the space
of fluid states and ${\cal S}_{\psi}$ the subspace of momentary
equilibrium fluid states.

The equivalence between this condition and the one assumed in
\cite{gl1} follows from the following argument. The map ${\cal F}_K$ is
injective so,
$
K_a M^a_0{}_{AB} \; \psi^B =0  \; \Rightarrow \; \psi^A =0
$
where $\psi^A = (\psi, \psi^a, 0)$. Due to the definition of indices
$A$ and $B$, the system of equations above represents a scalar
equation, a vector equation, and a symmetric two indices tensor
equation.

First, consider the scalar equation, the contraction of the vector
equation with $\zeta^a$, and the contraction of the two indices tensor
equation with $\zeta^a \zeta^b$. All this constitute a linear algebraic
system of three scalar equations for variables $K_a \psi^a$, $K_a
\zeta^a \psi$, and $2K_a \zeta^a \zeta^b \psi_b$. The injectivity
implies that the only solution for these three variables is zero and
so, the determinant of the coefficient matrix is different from zero.
Conversely, if the determinant of the coefficient matrix is different
from zero, then we conclude that $\psi =0$ and $\psi^a =0$ and then the
map ${\cal F}_K$ is injective.  By direct inspection it can be checked
that these coefficients are the same found in equations (41)-(43) in
\cite{gl1}.

Second, consider the vector equation and the contraction of the two
indices tensor equation with $\zeta^a$. This constitute a linear algebraic
system of two vector equations for the variables mentioned in paragraph
above, and $2 \zeta^a  K_{(a} \psi_{b)}$ and $K_b \psi$. Because
of injectivity, the only solution for all these variables is zero
and so, the determinant of the coefficient matrix is different from
zero. It can be checked, by direct inspection, that these coefficients
are the same found in equations (45)-(46) in \cite{gl1}.

Finally, consider the two indices tensor equation. It constitutes a
linear algebraic two indices equation for the variables mentioned in
the previous paragraph and $ K_{(a} \psi_{b)}$. Because of the injectivity,
the only solution for all these variables is zero and so, the
determinant of the coefficient matrix is different from zero. By direct
inspection it can be checked that these coefficients are the same found
in equations (48) in \cite{gl1}.  

\section*{\normalsize \sf APPENDIX B: GENERAL STABILITY THEOREM}

\setcounter{equation}{0}
\renewcommand{\theequation}{B\arabic{equation}}

Consider the Cauchy problem for a first order system of partial
differential equations,
\begin{eqnarray}
&&\frac{\partial v}{\partial t} = \Bigl(A^j_0 + \ep A^j_1(v, \ep)\Bigr)
\frac{\partial v}{\partial x^j} + \Bigl(B_0 + \ep B_1(v, \ep)\Bigr) v
\label{sistkor}\\
&&v(t=0) = f(x^j) \nonumber
\end{eqnarray}
where $v:\R \times {\bf T}^s \ar \R^n$, $A^j_1(v, \ep)$ and $B_1(v,
\ep)$ are smooth ($C^{\infty}$) functions of their arguments, and
$f(x):{\bf T}^s \ar \R^n$ is also smooth. Let $P$ denote the projector
in the kernel of $B_0$. For the solution $v$ of (\ref{sistkor}) we
define $v^{(0)} = P \hat v(0,t)$ and $w=v-v^{(0)}$.  The stability
theorem proved in  \cite{kor} states\cite{no-symmetry}.
\begin{theorem} \label{teokor}
Suppose that the matrices $A^j_0$, $B_0$, and $A^j_1$ are hermitian, and
the system (\ref{sistkor}) satisfy the ``relaxed stability eigenvalue
condition", ie., the following conditions hold.

(i) There is a constant $\delta >0$ such that the eigenvalues
$\lambda(k)$ of the symbol $\;i A^j_0 k_j + B_0$ satisfy $\re\{\lambda\}
\leq -\delta$ for all $k \in \Omega$, $k\neq0$.

(ii) The eigenvalues of $B_0$ satisfy either $\re\{\lambda(0)\}
\leq -\delta$ or $\lambda(0)=0$. 

(iii) ${\rm ker}\, B_0 \subset {\rm ker}\, B_1$.

\noindent Then, for $0\leq \ep\leq \ep_0$ with $\ep_0$ small enough,
the system (\ref{sistkor}) is a contraction for $w$ in a suitable norm,
equivalent to a Sobolev norm $H^p$ ($p> s+2$), and $v^{(0)}
\ar \const$ when $t\ar \infty$.
\end{theorem}
The statement in the above theorem that the system is contraction means
that there exists an $H$-norm, equivalent to the norm $H^p$, such that
$$
\frac{d}{dt}\|w\|^2_H \leq -(\delta + {\cal O}(\ep)) \|w\|^2_{H},
$$
this implies that, for $\ep$ small enough, there exists a global (in
time) smooth solution of the Cauchy problem (\ref{sistkor}) such that
there is an exponentially decaying bound for the Sobolev $H^p$ norm of
the $w$-part of its solution, and the $v^{(0)}$ part goes to a
constant when the time goes to infinity.

The only difference between the fluid equations (\ref{sistf2}) and
(\ref{sistkor}) is the presence of $N$ in front of the time
derivative.  This causes no difficulties, and the theorem above is
applicable by a simple redefinition of the scalar product used. With
this new scalar product, the eigenvalues problem for the fluid
equations becomes
$$
\det \left(i N^{-1}_0 N_0^j k_j + N^{-1}_0 I_0 - \lambda  E \right)=0.
$$
where $E$ is the identity matrix. Then, the conditions (i), (ii), and
(iii) of the theorem \ref{teokor} can be verified as is done in the
proof of theorem \ref{teo}.


\begin{thebibliography}{99}
\bibitem{equilibrium} It is known that strict equilibrium states are
more general, but they reduce to constant states when the space-time is
$\R \times \T^3$ and flat.
\bibitem{no-symmetry} A more general theorem without the condition of
symmetry is also proved in \cite{kor}. For the present application it
is enough to consider the symmetric case.
\end{thebibliography}
\end{document}